\documentclass[useAMS,usegraphicx,usenatbib]{mn2e}
\usepackage{aas_macros}
\usepackage[a4paper,centering, totalwidth=520pt, totalheight=700pt]{geometry}
\usepackage[fleqn]{amsmath} 
\usepackage{graphicx}
\usepackage{amssymb}
\usepackage{amsmath}
\usepackage{enumerate}
\usepackage{microtype}
\usepackage{color}

\usepackage{fixltx2e}[1999/12/01]

\newcommand{\kpc}{\ensuremath{\mathrm{kpc}}}
\newcommand{\Mpc}{\ensuremath{\mathrm{Mpc}}}

\newcommand{\vect}[1]{\boldsymbol{#1}}

\newcommand{\ii}{\mathrm{i}}
\newcommand{\ee}{\mathrm{e}}

\newcommand{\parder}[3][]{\frac{\partial^{#1} {#2}}{\partial {#3}^{#1}}}

\newcommand{\diff}[2][]{\mathrm{d}^{#1}{#2}}
\newcommand{\idiff}[2][]{\!\!\mathrm{d}^{#1}{#2}}

\newcommand{\kms}{\ensuremath{\mathrm{km\,s}^{-1}}}

\newcommand{\arcmint}{\ensuremath{\mathrm{arcmin}}}
\newcommand{\degt}{\ensuremath{\mathrm{deg}}}

\newcommand{\clight}{\ensuremath{\mathrm{c}}}

\newcommand{\vr}{\vect{r}}
\newcommand{\vtheta}{\vect{\theta}}
\newcommand{\vvartheta}{\vect{\vartheta}}

\newcommand{\gammat}{\gamma_{\mathrm{t}}}
\newcommand{\gammax}{\gamma_{\times}}

\newcommand{\deltaM}{\delta_{\mathrm{m}}}
\newcommand{\OmegaM}{\Omega_{\mathrm{m}}}

\newcommand{\fK}{f_K}

\newcommand{\zS}{z_{\mathrm{S}}}
\newcommand{\pS}{p_{\mathrm{S}}}
\newcommand{\chiS}{\chi_{\mathrm{S}}}
\newcommand{\fS}{f_{\mathrm{S}}}

\newcommand{\zL}{z_{\mathrm{L}}}

\newcommand{\chiL}{\chi_{\mathrm{L}}}
\newcommand{\fL}{f_{\mathrm{L}}}

\newcommand{\fSL}{f_{\mathrm{SL}}}

\newcommand{\zSi}[1]{z_{\mathrm{S},#1}}
\newcommand{\nSi}[1]{n_{\mathrm{S},#1}}
\newcommand{\chiSi}[1]{\chi_{\mathrm{S},#1}}

\newcommand{\EV}[1]{\left\langle{#1}\right\rangle}
\newcommand{\bEV}[1]{\bigl\langle{#1}\bigr\rangle}

\newcommand{\Interval}[1]{\mathbb{I}^{(#1)}}
\newcommand{\NIntervals}{N_\mathbb{I}}
\newcommand{\chiLlo}[1]{\chi_{\mathrm{L,hi}}^{(#1)}}
\newcommand{\chiLmid}[1]{\chi_{\mathrm{L}}^{(#1)}}
\newcommand{\chiLhi}[1]{\chi_{\mathrm{L,lo}}^{(#1)}}

\newcommand{\fLmid}[1]{f_{\mathrm{L}}^{(#1)}}
\newcommand{\zLmid}[1]{z_{\mathrm{L}}^{(#1)}}

\newcommand{\param}{\pi}
\newcommand{\data}{\xi}
\newcommand{\pred}{m}
\newcommand{\vdata}{\vect{\data}}
\newcommand{\vparam}{\vect{\param}}
\newcommand{\vpred}{\vect{\pred}}

\newcommand{\datacov}{\Sigma_\mathrm{d}}

\newcommand{\likelihood}{p}

\newcommand{\transposed}[1]{{#1}^{\mathrm{t}}}

\newcommand{\cfh}{CFHTLenS}

\title[{\cfh} data and cosmology-scaling of simulations]
{Cosmological constraints from the {\cfh} shear measurements using a new, accurate and flexible way of predicting nonlinear mass clustering.}

\begin{document}

\author[Angulo \& Hilbert]{
Raul E. Angulo$^{1}$ \& 
Stefan Hilbert$^{2,3}$. 
\\ $^{1}$Centro de Estudios de F\'isica del Cosmos de Arag\'on, Plaza San Juan 1,  Planta-2, 44001, Teruel, Spain.
\\ $^{2}$Exzellenzcluster Universe, Boltzmannstr. 2, 85748 Garching, Germany
\\ $^{3}$Max-Planck-Institut f{\"u}r Astrophysik, Karl-Schwarzschild-Str. 1, 85748 Garching, Germany.
}
\maketitle

\date{\today}

\begin{abstract}
We explore the cosmological constraints from cosmic shear using a new way of
modelling the non-linear matter correlation functions. The new formalism
extends the method of \cite{Angulo2010b}, which manipulates outputs of $N$-body
simulations to represent the three-dimensional nonlinear mass distribution in
different cosmological scenarios. We show that predictions from our approach
for shear two-point correlations at $1$ to $300$ arcmin separations are
accurate at the $\sim10$\% level, even for extreme changes in cosmology. For
moderate changes, with target cosmologies similar to that preferred by analyses
of recent Planck data, the accuracy is close to $\sim5$\%. We combine this
approach with a MonteCarlo Markov Chain sampler to explore constraints on a
$\Lambda$CDM model from the shear correlation functions measured in the
Canada-France Hawaii Telescope Lensing Survey ({\cfh}). We obtain constraints
on the parameter combination $\sigma_8 (\Omega_m/0.27)^{0.6} = 0.801 \pm
0.028$. Combined with results from CMB data, we obtain marginalised
constraints on $\sigma_8 = 0.81 \pm 0.01$ and $\Omega_m = 0.29 \pm 0.01$. These
results are fully compatible with previous analyses, which supports the
validity of our approach. We discuss the advantages of our method and the
potential it offers, including a path to incorporate in detail the effects of
baryons, among others effects, in future high-precision cosmological analyses.
\end{abstract}

\begin{keywords}
gravitational lensing: weak -- large-scale structure of the Universe -- cosmological parameters -- cosmology: observations -- cosmology: theory -- methods: numerical
\end{keywords}

\section{Introduction}

Gravitational lensing represents a great opportunity to explore fundamental
properties of our Universe, e.g. its mean matter density, the properties of
primordial density fluctuations, and the nature of dark energy \citep[see,
e.g.,][for a review]{Bartelmann2010}. Gravitational lensing shear in the
images of distant galaxies has been detected in many surveys \citep[early
results include][]{Tyson1990,Wittman2000,Bacon2000,Wilson2001,Refregier2002},
and it has been used to place direct constraints on cosmological parameters
\citep[e.g.][]{Hoekstra2002,Hoekstra2006,Fu2008,Schrabback2010,Semboloni2011,Jee2013,Huff2014}. 

One of the largest shear surveys carried out so far is the Canada-France Hawaii
Telescope Lensing Survey \citep[{\cfh},][]{Heymans2012}, which is based on
observations of the Canada-France-Hawaii Telescope
Legacy\footnote{\texttt{http://www.cfht.hawaii.edu/Science/CFHLS/}} (CFHTLS).
An analysis of the cosmological implications of the {\cfh} 2D shear correlation
function was presented in \cite{Kilbinger2013}. \cite{Fu2014} extended this
analysis by incorporating the information from higher-order correlation
functions. \cite{Kitching2014} focused on the cosmological constraints from the
reconstructed 3D mass distribution. 

The amount and accuracy of weak lensing data will be improved dramatically over
the next decade. Surveys such as the Dark Energy
Survey\footnote{\texttt{http://www.darkenergysurvey.org}} (DES),
J-PAS\footnote{\texttt{http://j-pas.org}} or
Euclid\footnote{\texttt{http://sci.esa.int/euclid}} will scan large fractions
of the sky with unprecedented detail, and are expected to reduce current
uncertainties in the shear measurements by an order of magnitude. The new data
will demand a new level of accuracy and sophistication in its analysis.

A crucial ingredient for a cosmological interpretation of shear data is a
description of the nonlinear mass distribution in the Universe as a function of
the underlying cosmological model. Analytic prescriptions based on perturbation
theory break down at the relevant scales, thus $N$-body simulations are the
most accurate way of providing such information \citep[see][for a recent
review]{Kuhlen2012}. Unfortunately, it is currently impossible computationally
to carry out thousands of simulations, as it would be required for such
analysis.  Thus, cosmic shear analyses have mostly relied on fitting functions
\citep[e.g.][]{Peacock1996} and/or the halo model
\citep[e.g.][]{Seljak2000,Smith2003}, even though these approaches exhibited
noticeable accuracy problems \citep[][]{White2004,Hilbert2009,Sato2009}.
Schemes known as emulators \citep{Habib2007,Heitmann2010,Schneider2011}, which
interpolate between a set of simulations in different cosmologies, have also
been proposed, but their application has been hampered by their limited
coverage of cosmological parameter space, nonlinear scales and redshift range
\citep[][]{Kilbinger2013}.  There is a large on-going effort by the community
aimed at improving the accuracy and scope of these approaches \citep[see,
e.g.,][]{Takahashi2012,Heitmann2014,Agarwal2014}.

In this paper we explore another way of modelling shear correlation functions,
which, we argue, provides a sufficiently flexible and precise theoretical
framework for analysing upcoming weak lensing surveys. Our method is based on
an extended version of the algorithm originally presented by \cite{Angulo2010b}
(thereafter AW10), which is able to predict the 3D mass distribution -- including
voids, filaments and self-bound structures -- expected in an arbitrary
cosmological model. In particular, the method provides predictions for the full
hierarchy of mass correlation functions, for halo mass functions and for the
cosmic velocity field.

The accuracy of the method is determined by the original $N$-body simulation
employed and by the distance in the cosmological parameter space between the
original and target cosmologies. Thus, the associated uncertainty can be
controlled by running a small number of $N$-body simulations with cosmological
parameters carefully chosen, and the realism of the predictions is that
inherited from the original simulation (which has been steadily increasing over
the last 30 years). A great advantage of our method is its full three
dimensionality.  Since cosmic structures are individually resolved, physically
motivated recipes (which can have an explicit dependence on halo mass or
assembly history) for how baryons redistribute mass inside halos
\citep[e.g.][]{vanDaalen2011,vanDaalen2014} can be applied. Also, non-trivial
survey geometries, selection functions and the source-lens clustering can all
be naturally incorporated in the analysis.

These extensions will be described in forthcoming publications. Here we focus
on demonstrating that accurate cosmological exploitation is possible using the
method. We will show that even with a single cosmology sampled, and with
current $N$-body simulations, it is possible to predict the signal measured by
{\cfh} with an accuracy of $\lesssim10\%$ (which decreases to $\sim5\%$ when a
smaller range of the cosmological parameter space is considered) and with
enough speed to explore almost $100,000$ cosmological models.

The outline of this paper is as follows. In Section 2 we provide the
theoretical background for the lensing quantities we will explore. Our model
for the shear correlation functions is described in Section 3. Section 4 is
devoted to test the performance of our model by comparing its predictions to
those from $N$-body simulations. In Section 5 we present the constraints on
cosmological parameters obtained from the {\cfh} data alone and combined with
CMB measurements. Finally, in Section 6 we present our conclusions and discuss
future developments of our formalism.

\section{Theory}

\subsection{Basic definitions}

Gravitational lensing, the deflection of light emitted from distant sources by
the gravity of matter structures along the line of sight, can shift and shear
the images of distant galaxies
\citep[e.g.][]{SchneiderKochanekWambsganss_book}. The observed image position
$\vect{\theta}=(\theta_1,\theta_2)$ of a source at redshift $\zS$ may thus
differ from the source's `true' angular position
$\vect{\beta}=\bigl(\beta_1(\vect{\theta},\zS),\beta_2(\vect{\theta},\zS)\bigr)$.
Image distortions caused by differential deflection can be quantified by the
distortion matrix 
\begin{equation}
\label{eq:lens_distortion}
 \left(\parder{\beta_i(\vect{\theta},\zS)}{\theta_j}\right)_{i,j=1,2} =
\begin{pmatrix}
 1 - \kappa - \gamma_1 & - \gamma_2 - \omega \\
 - \gamma_2 + \omega &  1 - \kappa + \gamma_1
\end{pmatrix},
\end{equation}
whose decomposition defines the convergence
$\kappa(\vect{\theta},\zS)$, the asymmetry $\omega(\vect{\theta},\zS)$, and the
complex shear $\gamma(\vect{\theta},\zS)= \gamma_1(\vect{\theta},\zS) + \ii
\gamma_2(\vect{\theta},\zS)$.

To first order (and to a good approximation in weak lensing); (i) the asymmetry
vanishes, (ii) the shear and convergence fields are related through simple phase
factors in harmonic space, and (iii) the convergence is given by a weighted
projection of the matter density along the line of sight:
\begin{equation}
\label{eq:convergence}
  \kappa(\vect{\theta}, \zS) =
	\frac{3 H_0^2\OmegaM}{2\clight^2} 
	\int_{0}^{\chiS}\idiff[]{\chiL}
	\bigl(1 + \zL\bigr)
	\frac{\fSL \fL}{\fS}
 \deltaM\big(\fL \vtheta, \chiL, \zL \big)
   .
\end{equation}
Here, $H_0$ denotes the Hubble constant, $\OmegaM$ the cosmic mean matter
density (in units of the critical density), $\clight$ the speed of light. We
use abbreviations $\zL=z(\chiL)$, $\chiL=\chi(\zL)$, $\chiS=\chi(\zS)$,
$\fL=\fK(\chiL)$, $\fS=\fK(\chiS)$, and $\fSL=\fK(\chiS-\chiL)$, where
$z(\chi)$ denotes the redshift and $\fK(\chi)$ the comoving angular diameter
distance for sources at comoving line-of-sight distance $\chi$,  and
$\fK(\chi)$ denotes the comoving angular diameter distance for comoving
line-of-sight distance $\chi$. Furthermore, $\deltaM(\fL \vtheta, \chiL, \zL)$
denotes the matter density contrast at comoving position $(\fL \vtheta, \chiL)$
and redshift $\zL$.

\subsection{Convergence correlations}

For a population of sources with redshift distribution $\pS(\zS)$, an effective
convergence can be expressed by
\begin{equation}
\begin{split}
\label{eq:effective_convergence}
  \kappa(\vect{\theta}) &= \int\idiff[]{\zS}\, \pS(\zS) \kappa(\vect{\theta}, \zS)
	\\&= 
	\int_0^{\infty}\!\!\diff{\chiL}
 g(\chiL)
 \deltaM\big(\fL \vtheta, \chiL, \zL\big) 
	,
\end{split}
\end{equation}
where $g(\chiL)$, the lensing weight factor, is defined as:
\begin{equation}
\label{eq:effective_convergence_lens_weight}
g(\chiL) =
\frac{3 H_0^2\OmegaM}{2\clight^2}(1+\zL)\fL
\int_{\zL}^{\infty}\!\!\diff{\zS}\, \pS(\zS)  \frac{\fSL}{\fS}
.
\end{equation}

Assuming a statistically homogeneous and isotropic universe, all two-point
correlation functions do not depend on spatial or angular positions, but only
on redshift and relative distances. Let angular brackets $\bEV{\dots}$ denote
the expectation for an ensemble of universes with a particular set of
cosmological parameters. Let
\begin{equation}
	\xi(|\vr|, z, z') \equiv \EV{\deltaM(\vr', z)\deltaM(\vr' + \vr, z')}
\end{equation}
denote the two-point correlation function of the matter density
contrast. Then, the convergence correlation
\begin{equation}
\label{eq:convergence_correlation_df}
	\xi_\kappa(|\vtheta|) = \EV{\kappa(\vtheta')\kappa(\vtheta' + \vtheta)}
\end{equation}
can be expressed as
\begin{multline}
\label{eq:convergence_correlation_full}
\xi_\kappa(\theta) = 
	\int_0^{\infty}\!\!\diff{\chiL}
	\int_0^{\infty}\!\!\diff{\chiL'}\,
 g(\chiL) g(\chiL')
\\\times
 \xi\left(\sqrt{\bigl(\fL + \fL'\bigr)^2\theta^2/4 + \bigl(\chiL'- \chiL\bigr)^2}, \zL, \zL'\right) 
.
\end{multline}

Define the projected correlation $w(r, z)$ of the
density contrast at redshift $z$ and projected comoving separation $r$ by
\begin{equation}
\label{eq:df_xi_deltaM_proj}
w(r, z) = \int_{-\infty}^{\infty}\idiff{\chi}\,\xi \bigl(\sqrt{r^2 + \chi^2}, z , z\bigr)
.
\end{equation}
Using a Limber-type approximation, the convergence correlation simplifies to
\begin{equation}
\label{eq:convergence_correlation_Limber}
\begin{split}
\xi_\kappa(\theta) &= 
	\int_0^{\infty}\!\!\diff{\chiL}
 g(\chiL)^2
w\bigl(\fL \theta, \zL \bigr)
.
\end{split}
\end{equation}

\subsection{Shear correlations}

In cosmic shear surveys, correlations of the shear field are the primary
observables. In analogy to the effective convergence, we define an effective
shear as:
\begin{equation}
\label{eq:effective_shear}
\begin{split}
  \gamma(\vect{\theta}) &= \int\idiff[]{\zS}\, \pS(\zS) \gamma(\vect{\theta}, \zS).
\end{split}
\end{equation}

The shear can be decomposed into a tangential component
$\gammat(\vtheta,\vvartheta)$ and a cross component
$\gammax(\vtheta,\vvartheta)$ with respect to any given direction $\vvartheta$
\citep[e.g.][]{SchneiderVanWaerbekeMellier2002}:
\begin{equation}
\label{eq:df_shear_tangential_and_cross_component}
\gammat(\vtheta,\vvartheta) + \ii \gammax(\vtheta,\vvartheta) = -\ee^{-2\ii\varphi(\vvartheta)} \gamma(\vtheta),
\end{equation}
where $\varphi(\vvartheta)$ denotes the polar angle of the vector
$\vvartheta$. With this decomposition, two translation and rotation invariant
shear correlation functions can be defined \citep[e.g.][]{Kaiser1992}:
\begin{equation}
\label{eq:xi_pm_definition}
  \xi_{\pm} \bigl(\lvert\vtheta \rvert\bigr) =
  \EV{\gammat(\vtheta',\vtheta)\gammat(\vtheta'+\vtheta,\vtheta)} \pm \EV{\gammax(\vtheta',\vtheta)\gammax(\vtheta'+\vtheta,\vtheta)}.
\end{equation}
These cosmic shear correlation functions can be directly estimated from the
observed image ellipticity of distant galaxies.

Exploiting the relation between shear and convergence in harmonic space, 
\begin{align}
  \xi_{+} \bigl(\theta \bigr) &= \xi_\kappa(\theta)
	\quad\text{and}\\
  \xi_{-} \bigl(\theta \bigr) &= \xi_\kappa(\theta) + \int_{0}^{\theta}\idiff{\theta'} \left(\frac{4\theta'}{\theta^2} - \frac{12 \theta^{\prime 3}}{\theta^4} \right) \xi_\kappa(\theta')
	.
\end{align}

Thus, by defining 
\begin{align}
  w_{+}(R, z) &= w(R, z)
  \quad\text{and}\\
  w_{-}(R, z) &= w(R, z)  + \int_{0}^{R}\idiff{R'} \left(\frac{4 R'}{R^2} - \frac{12 R^{\prime 3}}{R^4} \right) w(R', z)
,
\end{align}
we can write the shear correlations as
\begin{equation}
\label{eq:xi_pm_wl_integral}
  \xi_{\pm} (\vtheta) = 
		\int_0^{\infty}\!\!\diff{\chiL}
 g(\chiL)^2
w_{\pm}\bigl(\fL \theta, \zL \bigr)
.
\end{equation}
Therefore, observational measurements of shear correlations encode information
about the statistical properties of the mass in the Universe at different
cosmic epochs. 

\section{Data}
\label{sec:data}

\subsection{Shear correlation functions from {\cfh}}
\label{sec:data:shear}

We use the measurements of cosmic shear correlation functions, $\xi_{\pm}$ 
(c.f. Eq.~\ref{eq:xi_pm_definition}), by \citet{Kilbinger2013} 
from the Canada-France-Hawaii Telescope Lensing Survey \citep[{\cfh},][]{Heymans2012}. 
CFHTLenS is based on observations of the Canada-France-Hawaii Telescope Legacy 
(CFHTLS), and it currently offers one of the most informative astronomical
datasets for cosmic shear. The survey spans $154$ square degrees observed
in five optical bands $u^*$, $g'$, $r'$, $i'$, and $z'$ with a $5\sigma$
point-source apparent magnitude limit $i'_{AB} \sim 25.5$. 
The CFHTLenS analysis combines weak lensing data processing with THELI \citep{Erben2013},
shear measurement with lensfit \citep{Miller2013}, and photometric redshift measurement with PSF-matched photometry \citep{Hildebrandt2012}.
A full systematic error analysis of the shear measurements in combination with the photometric redshifts is presented in \citet{Heymans2012},
with additional error analyses of the photometric redshift measurements presented in \citet{Benjamin2013}.

\citet{Kilbinger2013} used approx. $75\%$ of the full CFHTLS area (selected
according to results from tests for residual systematic errors) to measure
shear correlation functions $\xi_\pm(\theta)$ on scales $0.9\,\arcmint \leq
\theta \leq 300\,\arcmint$. The background galaxies for the shear correlation
measurements have a mean number density $\bar{n} \sim 17\,\arcmint^{-2}$ and
mean redshift $\bar{z} \sim 0.75$. In this paper we will employ the
publicly-available source redshift distribution, measured shear correlations,
and data covariance\footnote{\texttt{http://www.cfhtlens.org}}.

\subsection{Cosmic Microwave Background measurements}
\label{sec:data:cmb}

We combine our cosmological constraints from the shear data with those
derived from Cosmic Microwave Background (CMB) measurements, which will help to
break degeneracies among cosmological parameters. Specifically, we use a
combination of the CMB temperature power spectrum from the 1st year data of the
Planck satellite \citep{PlanckCosmo2013}, CMB temperature measurements on small
angular scales from the Atacama Cosmology Telescope \citep[ACT,][]{Das2014} and
the South Pole Telescope \citep[SPT,][]{Story2013}, and polarization
measurements on large scales (low multipoles) from the Wilkinson Microwave
Anisotropy Probe \citep[WMAP,][]{Hinshaw2013}. For simplicity we refer to this
combination as ``Planck''.   

\section{Methodology}
\label{sec:methodology}

In this section we describe our model for the shear correlation expected from a
given lensing survey. In \S\ref{sec:scale} we describe an improved version of
the algorithm proposed by \cite{Angulo2010b}, which we use together with the
$N$-body simulations described in \S\ref{sec:nbody} to predict the projected
mass correlation functions in an arbitrary cosmology. In \S\ref{sec:discrete}
we describe how these can be combined to compute shear correlation functions.

\subsection{Rescaling mass correlations functions}
\label{sec:scale}

The first and most important ingredient to exploit shear observations is
predictions for the nonlinear mass correlation function as a function of
redshift for an arbitrary cosmological model. This is often given by fitting
functions, emulators or analytic prescriptions based on the halo model. Here,
we adopt a different path, exploiting the method of AW10, which alters the
length, mass and time units of an $N$-body simulation to mimic the
three-dimensional mass distribution of an $N$-body simulation carried out for a
different set of cosmological parameters. 

We refer to the original AW10 paper for details of the method, and to
\citet{Ruiz2011} and \citet{Mead2014} for additional tests. Here, we simply
recall that the method is able to rescale the mass distribution of a simulation
such that matter and halo power spectra are matched to better than $5$\% at all
redshifts (and $\lesssim 2\%$ deviation at $z=1$), over the whole range of
scales tested, i.e. $1$ to $500$ Mpc. In addition, the mass function of DM
halos is reproduced with an error $\lesssim10\%$.

Although the accuracy of the method of AW10 is already high, here we have
modified it slightly to enhance its performance. In the original formulation of
the method, the rescaling parameters are set by a match between the variance of
the linear mass field in the rescaled and target cosmologies. Here, we seek the
best scaling parameters by also requiring that the growth history be as close
as possible between the rescaled and target simulations. 

Explicitly, we compute the linear growth history for scaled cosmologies within
$\Delta \log a_{*} = 0.1$, where $a_*$ is the expansion factor in the original
cosmology that best represents the $a=1$ output in the target cosmology.  Then,
we select as a new scaled cosmology that with the closest growth history to
that of the target cosmology. We do this by comparing $10$ expansion factor
values (uniformly-spaced between $a=0.1$ and $1$). We note that the weight
given to matching the growth and the linear variance is, at the moment, empirical.
However, in principle it is possible to adapt the weight for each target cosmology 
in order to minimize a suitably defined global uncertainty.

This modification aims at taking into account the assembly history of dark
matter halos, which is expected to largely determine the density profile of
halos \citep{Ludlow2013,Ludlow2014}. Thus, this is expected to increase the
accuracy of the AW10 method on small scales. In \S\ref{sec:nbody-scale} we
will present a comparison of the original and modified AW10 algorithm
applied to predict {\cfh} shear correlations.

The final step in the AW10 algorithm is to perform a quasi-linear distortion of
the 3D mass distribution that accounts for differences in the shape of the
primordial power spectrum between the scaled and target cosmologies. We will
refer to this as large-scale structure correction.

The algorithm described in this section opens the remarkable option of creating
3D maps for the expected mass distribution in any cosmology. We simply need an
$N$-body simulation with enough volume, mass and time resolution to match the
requirements placed by the dataset to be modelled. In the next section we
describe a suite of simulations that meets these requirements for analysing the
{\cfh} data.

\subsection{$N$-body simulations} 
\label{sec:nbody}

Our analysis uses the projected mass correlation functions measured in the
Millennium Simulation (MS) suite: MS-I \citep[][]{Springel2005a}, MS-II
\citep{Boylan-Kolchin2011} and MXXL \citep{Angulo2012a}. The MS runs adopt a
matter density of $\Omega_\mathrm{m}=0.25$ in units of the critical density, a
cosmological constant with $\Omega_\Lambda=0.75$, a Hubble constant $h=0.73$ in
units of $100\, \mathrm{km}\,\mathrm{s}^{-1}\Mpc^{-1}$, a spectral index
$n_s=1$ and a normalisation parameter $\sigma_8=0.9$ for the primordial linear
density power spectrum. In Table~\ref{TabSimParam} we summarize the box size,
particle mass and the Plummer-equivalent gravitational force softening employed
in each run of the MS suite.

\begin{center}
 \begin{table}
   \caption{Properties of the simulations used in this work. N$_{\rm p}$ is the number of
     particles; $L_{\rm box}$ is the box size of the simulation; $\epsilon$ is softening 
     length and $m_{\rm p}$ is the particles mass. }
   \begin{tabular}{l c c c c c c c}\hline \hline
     Simulation     & N$_p$               & $L_{\rm box}$ & $\epsilon$ & $m_{\rm p}$             & \\
                    &                     & [Mpc$/h$]     & [kpc$/h$]
                    & [M$_\odot/h$] & \\ 
\hline
     MS-XXL         &  6720$^3$           & 3000 & 10    & 6.17$\times     10^9$ & \\ 
     MS-I           &  2160$^3$           & 500  & 5     & 8.61$\times 10^8$ & \\
     MS-II          &  2160$^3$           & 100  & 1     & 6.89$\times 10^6$ & \\
     Test-Small     &  1080$^3$           & 250  & 5     & 8.61$\times 10^8 \,  \left(\frac{\Omega_{\rm m}}{0.25}\right)^{1/3}$ & \\    
     Test-Large     &  512$^3$           & 1500 & 60    & 1.75$\times 10^{11} \, \left(\frac{\Omega_{\rm m}}{0.25}\right)^{1/3}$ & \\      
\hline
   \end{tabular}
   \label{TabSimParam}
 \end{table}
\end{center}

The cosmological parameters of the MS are different to those preferred by
recent analyses, in particular, the value of $\sigma_8$ is about 10\% higher
than current best fit values \citep[e.g.][]{PlanckCosmo2013}. This is, however,
an advantage for our analysis since higher fluctuation amplitudes allow the MS
to be scaled to a wider range of cosmologies. To further increase the scalable
region of the parameter space, we have evolved the full MS-I beyond $z=0$,
until $z=-0.846$ (or, equivalently to an expansion factor of $6.5$) using the
memory efficient version of the {\tt L-Gadget3} code described in
\cite{Angulo2012a}. We store $20$ additional snapshots, roughly equally spaced
in linear growth factor. This allow us to predict the cosmic structure in, for
instance, a cosmology with a value of $\sigma_8$ equal to $1.1$.

\subsection{Projected correlation functions} 
\label{sec:wrp}

At each output time, we compute the projected correlation 
function of the mass distribution as follows:
\begin{equation}
w(r, z) = 2 \int_{x_{\rm min}}^{x_{\rm max}} {\rm d}x \, \xi \bigl(\sqrt{x^2 + r^2}, z\bigr)
\label{eq:wrp}
\end{equation}
where $\xi(r, z)$ denotes the 3D correlation function at separation $r$ of the matter density contrast at redshift $z$.
We compute $\xi$
in Fourier space employing a $4096^3$ grid, following the procedure described
in \cite{Angulo2014}. We repeat the calculation twice after folding the box
size $8$ times in each direction, which allows us to compute $\xi(r)$ on much
smaller scales: for the MS-I this scale is $r_{\rm min} = 500/8/8/4096 =
2\,\kpc/h$. In the case of the MXXL, only the density field on a $2048^3$ grid
was stored at every output, so we use these when computing the correlation
function. We set $x_{\rm min} = 10^{-4}\,\Mpc/h$ and $x_{\rm max} =
100\,\Mpc/h$. We also set $\xi(r) = \xi(r_{\rm min})$ for all $r < r_{\rm
min}$, since  $x_{\rm min} < r_{\rm min}$.


\begin{figure}
\begin{center}
\includegraphics[width=\linewidth]{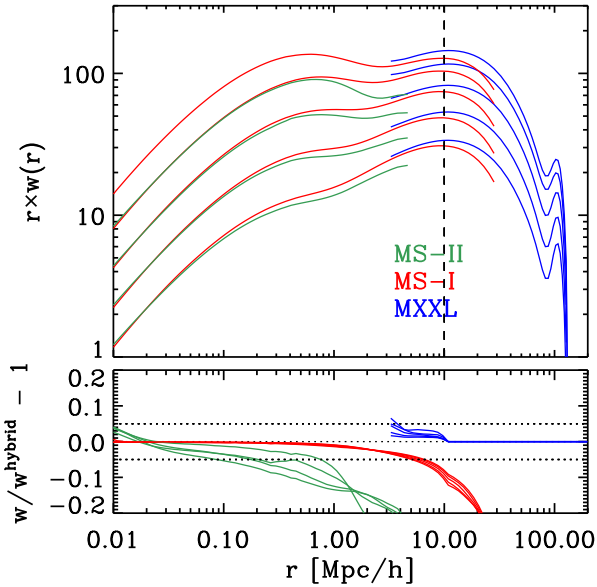}
\caption{
\label{fig:wrp} 
The projected mass correlation function, $w(r)$, measured in the
Millennium Simulations suite, at 5 different output times. From
bottom to top these are $z=1.5$, $0.9$, $0.4$, $0.0$, and $-0.25$.
For each redshift, coloured lines show results for the MS-II (green), 
MS (red) and MXXL (blue) runs, as indicated by the figure legend. 
Note we display $r\times w(r)$ to enhance the effective range shown.
}
\end{center}
\end{figure}

For testing the correctness and accuracy of our calculations, we also
compute the projected correlations in a different way: For each simulation output,
the matter distribution is first projected onto the simulation box faces. Then
the correlation of the projected matter densities is computed using a 2D Fast
Fourier method. To compute the correlations on large scales, coarse meshes
covering the whole box faces are used. For the correlation on smaller scales,
the box faces are tiled with a set of smaller but finer meshes, on which the
correlations are computed using appropriate zero-padding to account for
non-periodic boundary conditions \citep[][]{Hilbert2011}. A comparison of the
results from this approach to the first approach based on 3D correlations shows
a good agreement. For scales $r<50\,\Mpc/h$, both estimates agree at the 3\%
level for the $z=0$ output of the MXXL simulation. On larger scales, cosmic
variance dominates the correlation function of the projected mass field. 

The identical output times and cosmological parameters of the MS runs enable a
detailed comparison as well as a joint use of these simulations. In
Fig.~\ref{fig:wrp} we show the projected correlation functions measured in the
three MS at $5$ output times: $z=1.5$, $0.9$, $0.4$, $0.0$, and $-0.25$, from
bottom to top. By comparing the three simulations we can clearly see the
effects of force resolution and box size, both of which decrease the amplitude
of the projected correlation function. The finite box size and lack of long
wavelengths have a greater impact on the projected correlations than on
spherically averaged correlation functions.  The finite-volume effect is found
$10$\% at a scale of $1/50$ of the box size, roughly independent on redshift,
which is in good agreement with analytical estimates.  The effect of the force
resolution also appears to be redshift independent and of a magnitude of 10\%
at a scale twice as large as the softening length ($5\,\kpc/h$ in the MS-I). 

We see that none of the simulations individually can predict the projected
correlation function accurately, thus, we have built a hybrid correlation
function, $\xi^{\rm hybrid}$, by combining the results of the MXXL on scales
larger than $r_{\rm t} = 7\Mpc/h$, and those of the MS-I on smaller scales.
Explicitly:

\begin{eqnarray}
 \xi^{\rm hybrid}(r,z) &=&  \xi^{\rm MS-I}(r,z)\, {\rm erfc}\left(\frac{r-r_{\rm t}}{2.5}\right)/2  + \\ 
                       & &  \xi^{\rm MXXL}(r,z) \left[1 - \frac{1}{2}{\rm erfc}\left(\frac{r-r_{\rm t}}{2.5}\right) \right]
\end{eqnarray}

For $z<0$, we use the MS-I evolved into the future (c.f. \S\ref{sec:nbody}) and
the $z=0$ MXXL measurement linearly scaled to the appropriate (negative)
redshift.  Although not shown here, we note that, at the transition scale
($7\,\Mpc/h$), the measurements from the MS-I and MS-XXL agree at the $2-3\%$
level, over all redshifts examined. At this point, we refrain from using the
MS-II, since the MS-I is sufficient to model the smallest scales relevant
for our purposes.

The bottom panel of Fig.~\ref{fig:wrp} shows the
fractional differences of the projected correlations of the individual simulations
relative to the hybrid result. At all
redshifts, we can see that our combined model employs different simulations where they
are most accurate, resulting in an accuracy of $5$\% or better. We use
these combined correlations together with the scaling method described above for the
remainder of this paper. However, we note that future simulations with higher
force resolution and larger volumes should provide an even more accurate
starting point for our formalism.

\subsection{Modelling the shear correlation function measurements}
\label{sec:discrete}

After computing the hybrid projected correlation functions described in above,
we find the scaling parameters for a given target cosmological model. Then we
change the length unit of the projected correlation functions accordingly, and
re-associate each output with a different cosmic time.

For computing the shear correlations, we first partition the lightcone into
intervals $\Interval{k}=\bigl[\chiLlo{k},\chiLhi{k}\bigr)$,
$k=1,\ldots,\NIntervals$ that match the scaled output times of our $N$-body
results.  Within each $\Interval{k}$, we select a point
$\chiLmid{k}\in\Interval{k}$ by choosing the comoving line-of-sight distance to
the scaled snapshot time in the target cosmology. Then, we split the integral
\eqref{eq:xi_pm_wl_integral} into integrals over the intervals, and approximate
the integrands by their mid values:

\begin{equation}
\label{eq:xi_integral_to_sum}
\begin{split}
  \xi_{\pm} (\vtheta) &=
	  \sum_{k}
     \int_{\chiLlo{k}}^{\chiLhi{k}}\!\!\diff{\chiL}
     g\bigl(\chiL\bigr)^2
	   w_{\pm}\bigl(\fL \theta, \zL \bigr)
  \\&\approx
	  \sum_{k}
     \left( \chiLhi{k} - \chiLlo{k} \right)
     g\bigl(\chiLmid{k}\bigr)^2
     w_{ \pm}\bigl(\fLmid{k} \theta, \zLmid{k} \bigr)
		.
\end{split}
\end{equation}

The discretisation allows us to compute the shear correlations from the matter
correlations at a finite number of redshifts, in particular those of snapshot
outputs of a given simulation. This approximation becomes inaccurate if the
simulation output times sample the matter evolution to sparsely. Discretisation
accuracy tests will be discussed in  \S\ref{sec:ray}.

As an alternative to the sum \eqref{eq:xi_integral_to_sum}, one may obtain a
better approximation by using the matter correlations of the snapshots to
construct an interpolation of the matter correlation as function of scale and
redshift, and then feeding that interpolation function into the integral. Here,
we will not pursue this option since, as discussed later, our model already
appears accurate enough to exploit the {\cfh} measurements.

In this work, we use a source galaxy redshift distribution that is given as
galaxy numbers $\nSi{i}$ at a number of discrete source refshifts $\zSi{i}$
(i.e. as a source redshift histogram). In that case, the lensing weight factors
\eqref{eq:effective_convergence_lens_weight} can be expressed as sums, too,
\begin{equation}
\label{eq:effective_convergence_lens_weight_sum}
g(\chiL) =
\frac{3 H_0^2\OmegaM}{2\clight^2}(1+\zL)\fK{\chiL}
\sum_{i}\nSi{i}  \frac{\fK(\chiSi{i} - \chiL)}{\fK(\chiSi{i})}.
\end{equation}

\section{Tests}
\label{sec:tests}

We now test the methodology presented in the previous section. We start by
comparing our predictions for shear correlation functions using the discretised integrals 
to the predictions from the continuous Limber and full integrals (\S\ref{sec:limber}). 
Next, we compare our predictions to results from full ray-tracing (\S\ref{sec:ray}).
We then asses the performance of the cosmology-scaling method by carrying out
direct $N$-body simulations covering a wide range of target cosmologies
(\S\ref{sec:nbody-scale}).

\subsection{Limber approximation and discretised integrals} 
\label{sec:limber}

\begin{figure} 
\includegraphics[width=\linewidth]{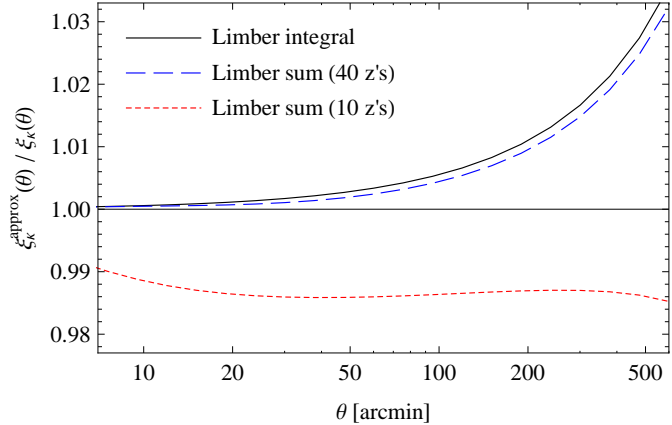}
\caption{
\label{fig:Limber_error} 
Systematic error on the computed convergence correlation $\xi_\kappa(\vtheta)$ to sources at redshift $z=0.75$ as a function of pair separation $\vtheta$ due to Limber-type approximations. Shown is the ratio of the approximated expression and the full expression \eqref{eq:convergence_correlation_full} for the Limber integral \eqref{eq:convergence_correlation_Limber} (black solid line), and for its discretisation \eqref{eq:xi_integral_to_sum} using either 40 (blue dashed line) or just 10 redshifts (red dotted line).
}
\end{figure}

The predictions in our approach are based on a Limber-type approximation \eqref{eq:convergence_correlation_Limber} for the convergence correlation, and further discretisation \eqref{eq:xi_integral_to_sum} of the resulting integral. To estimate the effect of these approximations, we first model the matter density correlation by a simple fit to the measured correlation:
\begin{equation}
	\xi(r,z,z') =  \frac{1}{1 + r^2 / r_\mathrm{s}^2}\frac{1}{1 + r^2 / r_\mathrm{c}^2}\exp\left(-\frac{\chi(z) + \chi(z')}{\chi_\mathrm{c}} \right)
.
\end{equation}
With the parameters $r_\mathrm{s} = 0.1 h^{-1} \Mpc$, $r_\mathrm{c} = 100 h^{-1} \Mpc$, and  $\chi_\mathrm{c} = 3200 h^{-1} \Mpc$, the fit captures the scale and redshift dependence of the correlation well enough for our purposes. We then use this fit as input matter correlation in the expressions \eqref{eq:convergence_correlation_full}, \eqref{eq:convergence_correlation_Limber} and \eqref{eq:xi_integral_to_sum} for the convergence correlations.

The results of this test are illustrated in Fig.~\ref{fig:Limber_error}. The errors introduced by the Limber approximation are well below 1\% for separations $\theta < 3\,\degt$. And even for scales $\theta\sim 5\,\degt$, the errors stay below 5\%. The discretisation errors are very small (and of opposite sign to the Limber approximation error), too, e.g. $1-5\%$ even if only $\sim10$ redshifts are used in the sum (for the more typical 40 redshifts, the error is much smaller). 

\subsection{Discretised integrals versus ray-tracing simulations} 
\label{sec:ray}

\begin{figure} 
\includegraphics[width=\linewidth]{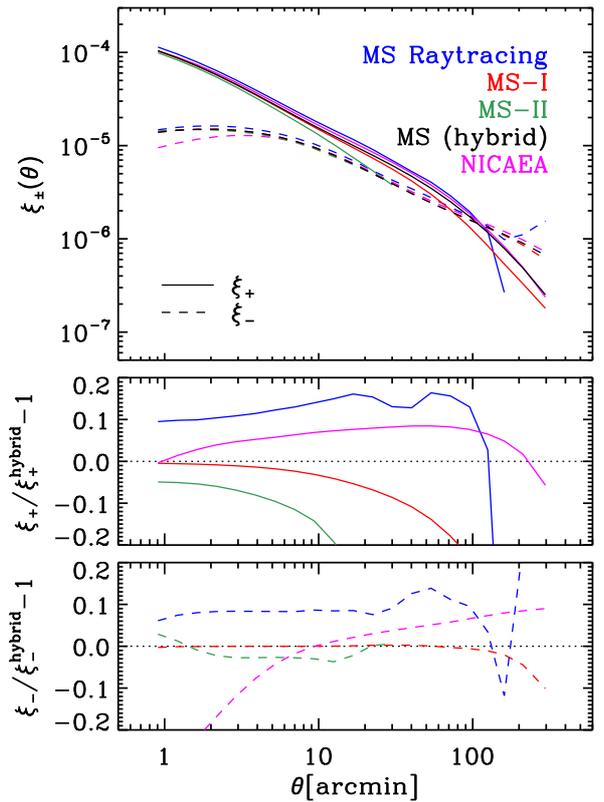}
\caption{
\label{fig:ray} 
Comparison of the shear correlation functions predicted by several
methods: a full ray-tracing algorithm applied to the MS-I simulation
(blue lines), a halo-model as implemented by the {\tt NICAEA} code
(magenta lines), a discretised integrals (c.f. Eq. 18) employing
a hybrid correlation function (our preferred model, black lines), or
those measured directly in the MS-I simulation (red lines) or
 MS-II (green lines). The middle and bottom panels show the fractional
 differences with respect to our preferred model for $\xi_+$ and $\xi_-$,
 respectively.}
\end{figure}

In Fig. \ref{fig:ray}, we present a comparison of the {\cfh} shear correlation
functions predicted for the MS cosmology by (i) a full ray-tracing simulation
performed on the MS-I (blue lines), (ii) a commonly used method based on the
halo model (magenta lines), and (iii) the algorithm described in the previous
section. For the latter we display cases where the input mass correlation
function was computed from the MS (red lines), the MS-II (green lines), and our
fiducial model based on the hybrid correlation functions (black lines)
described in \S \ref{sec:wrp}. In the middle and bottom panels, we show
fractional differences of these models with respect to this hybrid one. In all
cases, we employ the redshift distribution of background galaxies provided by
the {\cfh} collaboration.

Firstly, it is clear that the MS-I and MS-II simulation boxes are not big
enough to accurately describe correlations on scales larger than $\sim30$ and 3
arcmin, respectively. As it was also shown in Fig. \ref{fig:wrp}, the lack of
long-wave modes reduces the amount of correlation on large scales.  This
highlights the need of the MXXL simulation for our analysis, and for a correct
interpretation of shear signals. By comparing our hybrid model with the
predictions of the MS-II simulation, we can see that the much higher force
resolution of the latter simulation is important only at the $5$\% level and
for the smallest scales, which justifies our choice of not including it in our
hybrid predictions. We note that these scales might also be contaminated by
baryonic effects, which we do not attempt to model here.

We now compare our predictions with the results of the ray-tracing algorithm
(blue lines) described in \cite{Hilbert2009}. In brief, this method constructs
multiple-lens-plane lightcones from the MS outputs, and then traces light rays
backward through this series of lens planes, computing light deflections at the
lens planes from the projected matter distribution on these planes.  We show
correlation functions estimated from 64 ray-traced fields of $4 \times
4\,\degt^2$. Error bars display the standard deviation over them. We note that
multiple realisations are needed to perform a more accurate comparison: Since
the ray-tracing procedure involves a light-cone construction, the predictions
are more affected by cosmic variance than our discretised integrals (which use
the full simulation box at each output time, instead of a small cone segment). 

We can see that these results agree reasonably well with our (comparatively)
very simplified treatment of gravitational lensing effects. This supports the
validity of our approximations and implementation. There is, however, a
systematic excess in the ray-tracing correlations of about $10-15\%$ for both
$\xi_+$ and $\xi_-$ with respect to our model. The discrepancy increases to
about a factor of two when compared to the MS-I based predictions on large
scales. We have investigated several possible sources for the discrepancy.
However, we have found none apart from cosmic variance that could plausibly
explain this finding.\footnote{ The Limber approximation, for example, causes
errors $\lesssim1\%$ on the scales of interest.} Since this systematic
variation is not significant compared to the uncertainty in the {\cfh} data,
we postpone a deeper exploration of this issue to a future publication.

Finally, we compare our method with predictions based on the halo model -- one
of the most popular models employed in the literature. In particular, magenta
lines display the predictions of the implementation in \textsc{nicaea}
\citep{Nicaea2012}. Similarly to the ray-tracing case, we find a reasonably
good agreement with our model, with a discrepancy of $< 10\%$ for $\xi_+$. This
model, however, performs poorly on small scales for $\xi_-$, where
underestimates the signal by a factor of $2$, with respect to the raytracing
results and those of our model. We note that recent improvements to the halo
model \citep[e.g.][]{Takahashi2012} should produce more accurate predictions.

\begin{figure*} 
\includegraphics[width=0.85\linewidth]{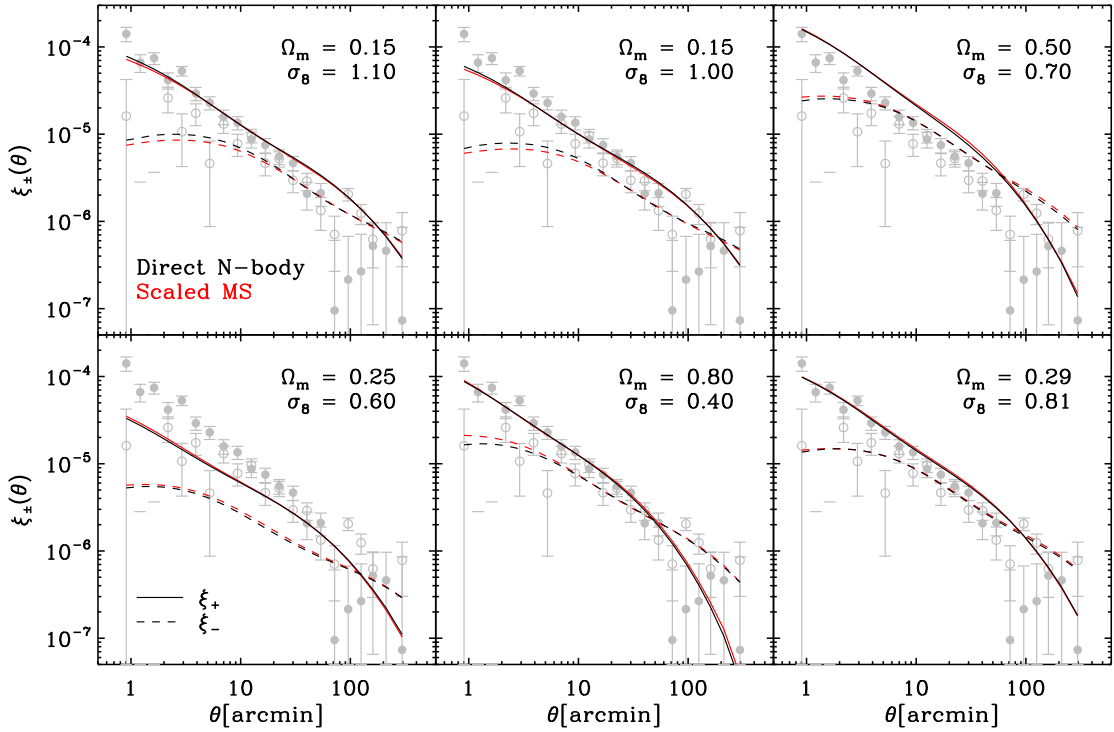}
\caption{
\label{fig:test} 
Comparison between the predictions of the theoretical model described in
\S\ref{sec:scale} (red lines) and direct $N$-body simulations (black lines, see
\S\ref{sec:nbody}). Each subpanel displays results for $\xi_{+}(\theta)$ (solid
lines) and $\xi_{-}(\theta)$ (dashed lines) for a different combination of
$\Omega_\mathrm{m}$ and $\sigma_8$, as indicated by the figure legends.  {\cfh}
measurements together with their uncertainty (grey symbols and error bars) are
shown for comparison. 
}
\end{figure*}

\begin{figure*} 
\includegraphics[width=0.85\linewidth]{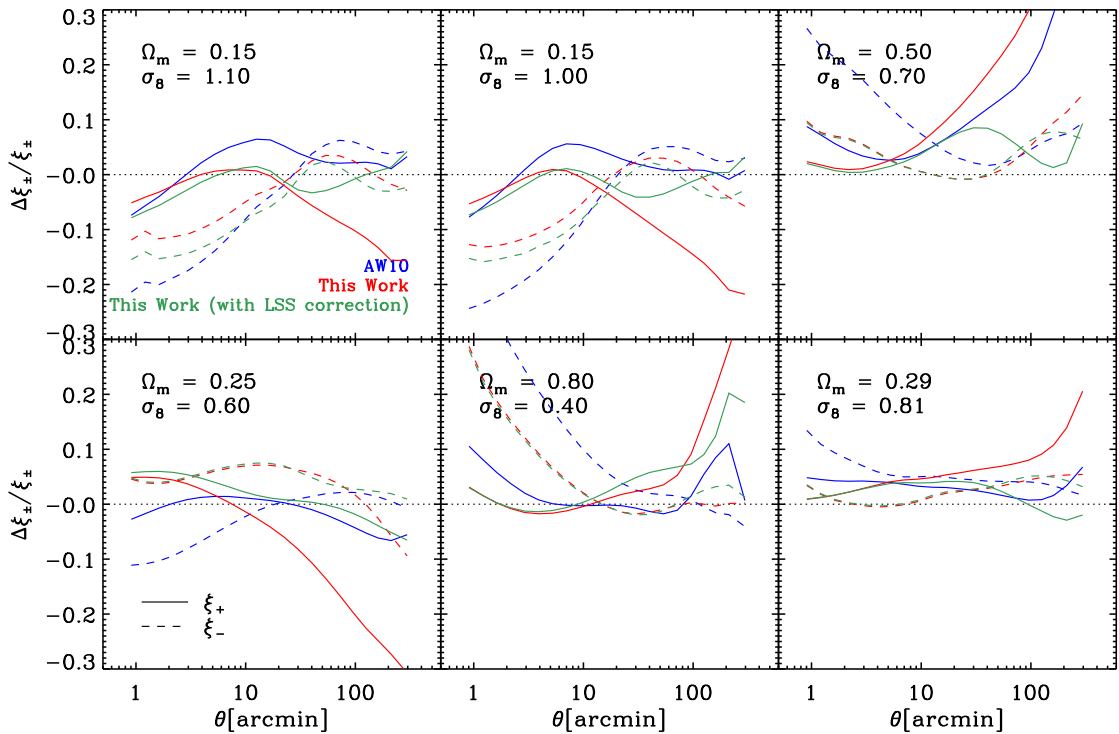}
\caption{
\label{fig:test2} 
Fractional difference between the shear correlation functions, $\xi_{\pm}$,
predicted by direct $N$-body simulations and three slightly different
cosmology scaling methods.  
}
\end{figure*}

\subsection{Cosmology scaling versus direct $N$-body simulations}
\label{sec:nbody-scale}

We now probe the accuracy of the scaling method described in \S\ref{sec:scale}
in predicting $\xi_{\pm}$ as a function of cosmology. For this, we have carried
out $6$ $N$-body simulations employing different combinations for $\Omega_m$
and $\sigma_8$. Specifically, we use: $(\Omega_m, \sigma_8)$ = $(0.15,1.1)$,
$(0.15,1.0)$, $(0.25, 0.5)$, $(0.5, 0.7)$, $(0.8, 0.4)$, $(0.29,0.81)$, which
were chosen to approximately cover the whole region of the parameter space we
will explore.  For each cosmology we have run two different simulation boxes,
$L_{\rm box}=250\,\Mpc/h$ and $1500\,\Mpc/h$, referred to as Test-Small and
Test-Big. The numerical parameters of these simulations are given in Table
\ref{TabSimParam}. We highlight that the output structure is identical to those
of the MS, and that the mass and force resolution of the  Test-Small suite
matches that of the MS-I run. In an analogous manner to the construction of our
fiducial model, we combine the measured correlation functions of the Test-Small
and Test-Big simulation suites to accurately cover the whole relevant range of
scales.

Fig. \ref{fig:test} shows one of the main results of this paper: a comparison
of the shear correlation functions predicted from direct simulations and from
our algorithm. In all cases, our procedure correctly captures the dependence of
$\xi_{\pm}$ on cosmology. The agreement between the direct and scaled
predictions is remarkable, and in several cases and scales, both predictions
are indistinguishable by eye! The precision is particularly high for $\xi_{+}$
where differences are always smaller than $10\%$, and somewhat lower for
$\xi_{-}$, specially on small angular scales where the differences can reach up
to $17\%$. We emphasise that we are considering very extreme cosmologies.  When
we focus on a cosmology compatible with the latest Planck results (bottom right
panel), the differences are $<3-4\%$ for both $\xi_{+}$ and $\xi_{-}$.

We have a closer look at the error in our procedure in Fig.~\ref{fig:test2},
where we display the fractional uncertainty introduced by our scaling
procedure.  Specifically, we plot $\xi_{\pm}^{\rm scaled}/\xi_{\pm}^{\rm
N-body} - 1$, where $\xi_{\pm}^{\rm scaled}$ is one of three cases: our
cosmology-scaling algorithm with and without the large-scale correction (green
and red lines, respectively), and the original AW10 without the large-scale
correction (blue lines).

Firstly, we note that there is a pattern in the sign of the deviations for all
three methods. Cosmologies that require a length scaling larger than the unity
have positive deviations, whereas those with a scaling scaling smaller than
unity have negative deviations. As the smallest angular scales predicted by the
MS are affected by its limited force resolution, this suggests that force
resolution is responsible for at least some of the discrepancy on small scales
shown in Fig.~\ref{fig:test2}.

By comparing red and blue lines we can see that, for all cosmologies, our
modified AW10 method performs better than the original one on small scales, in
some cases reducing the error by a factor of $2$ and confirming our
expectations set in \S~\ref{sec:scale}. More similar growth histories imply
more similar formation times and concentration-mass relation, which in turn
impacts positively the accuracy of the scaled correlation functions on small
scales. On larger scales, the original method performs better. As it does not
have the additional constraint of matching growth histories, it is a better fit
to the overall shape of the linear theory power spectrum. Nevertheless, this is
in principle not important since the large-scale correction brings back the
high accuracy of the method at all scales. 

It is also clear that the larger the difference in cosmological parameters, the
less accurate the scaled predictions are. Below 100 arcmin, for $\xi_+$, the
largest fractional difference is found for the $(0.5,0.7)$ cosmology. Due to
the high value of $\Omega_m$, this simulation does not feature a slow down of
structure formation caused by a cosmological constant, as it is the case for
the MS cosmology. On the other hand, the $(0.81,0.29)$ cosmology shows the
smallest fractional difference; it has a similar power spectrum and growth
history as the MS. This is an interesting feature since it opens up the
possibility of suitably defining locations in the cosmological parameter space
for performing direct $N$-body simulations designed to be scaled.  Such an
ensemble of simulations can then ensure a given accuracy of scaled shear
correlations.

As we have seen here (and also discussed in AW10), the large-scale correction
is an important aspect of the method. Unfortunately, the computational cost
associated is too high for it to be used in an exploration of a cosmological
parameter space. Although it is possible to perform a fast large-scale correction
directly on the measured correlations using low-pass-filtered 
linear-theory correlation functions, at all scales and for all cosmologies, even the
methods without this correction have a small associated systematic error: They
range from $20$ to $60$\% of the square root of the diagonal elements of the
data covariance matrix at the respective scale. This implies that the cosmology
scaling without the large-scale correction is indeed accurate enough for the
exploitation of {\cfh} data, and thus, we will use this for the remainder of
our paper. We note we choose our modified AW10 scheme over the original one
because large angular scales in the data show the largest variance. Thus,
relative to the uncertainties in the measurements, accuracy on large scales are
less important than that on smaller angular scales. 

To summarize this section, we conclude that the algorithm proposed and
developed here is of high accuracy for the cosmological exploitation of the
{\cfh} data. At each point of the parameter space would be possible to quickly
predict the full three-dimensional and nonlinear structures expected in any
cosmological model, together with the expected lensing shear they produce.

\section{Cosmological constraints}

We now apply the new algorithm presented and tested in previous sections to
obtain cosmological constraints from the {\cfh} data. In \S\ref{sec:pars} we
discuss the cosmological parameters to be explored, and also describe the
method used to sample their probability distribution function. In
\S\ref{sec:constraints} we present and discuss the cosmological implications of
our analysis.

\subsection{Parameter Space and Likelihood calculation} \label{sec:pars}

In this paper we explore the parameters of the minimal $\Lambda$CDM model,
where we assume Gaussian and adiabatic primordial fluctuations, a flat space
geometry, a dark energy component in the form of a cosmological constant, and a
negligible contribution of massive neutrinos and tensor perturbations. This
simple model is specified by $5$ parameters:

\begin{equation}
\vect{\pi} = (\sigma_8, \Omega_m, \Omega_b, n_s, h), 
\end{equation}

\noindent the $rms$ linear density fluctuations in spheres of radius $8\Mpc$,
$\sigma_8$; the total mass density in units of the critical density, $\Omega_m
= \Omega_{\rm dm} + \Omega_{b}$,  where $\Omega_{b}$ and $\Omega_{\rm dm}$
refer to baryons and dark matter, respectively; the primordial spectral index,
$n_s$; and the present-day Hubble constant, in units of $100\,\kms\,{\rm
Mpc^{-1}}$. The priors for these parameters are assumed to be flat and over the
range: $\sigma_8 \in
[0.4,1.1]$, $\Omega_m \in [0.1,0.8]$, $\Omega_b \in [0,0.1]$, $n_s \in
[0.7,1.2]$ and $h \in [0.4,1.0]$.

We assume that the probability of observing a given set of cosmic shear values
$\vdata=\bigl(\xi_+(\vartheta_1),\ldots\bigr)$ is given by a multivariate
normal distribution

\begin{equation}
\label{eq:like}
\likelihood (\vdata | \vparam) 
\propto \exp \left\{-\frac{1}{2} \transposed{\left[\vdata - \vpred(\vparam) \right]} \datacov^{-1} \left[\vdata - \vpred(\vparam) \right] \right\}.
\end{equation}

Here, $\datacov$ is the data covariance matrix as computed by
\cite{Kilbinger2013} for the {\cfh} shear correlation functions.
$\vpred(\vparam)$ denotes the expected cosmic shear signal computed with our
model described in \S\ref{sec:methodology}. We recall that this model shows an
accuracy better than 10\% over the whole range of scales explored even for very
extreme cosmologies and without any nuisance parameters. When
considering target cosmologies closer to the best fit values of the
$\Lambda$CDM model, the accuracy of our model increases to the few percent
level. 

In our model we neglect the effects induced by ``baryonic physics'', which can
redistribute mass inside halos through gas cooling and feedback from star formation, supernovea, and AGNs possibly affecting
the shear signal \citep[e.g.][]{Semboloni2011,Zentner2013}. However, as shown by \cite{Kilbinger2013}, excluding small
scales ($\theta < 17\,\arcmint$ or $\theta < 53\,\arcmint$) does not
significantly shift the cosmological constraints derived, which suggests that
the effect is statistically unimportant given the accuracy of the {\cfh}
measurements.

Nevertheless, we note that due to the 3D nature of our models, where individual
halos are resolved for each target cosmological model, such effect could in
principle be parametrized and incorporated into our formalism. The additional
uncertainty can be naturally marginalised over, providing robust cosmological
constraints. Carrying out such analysis, although possible, is still a tough
computational challenge. However, continuous development in
algorithms and supercomputing power make us optimistic about a successful
outcome in the near future.

We sample the posterior probability using our own implementation of the
Markov Chain Monte Carlo algorithm (MCMC).  At each step of the chain, we use
\textsc{CAMB} \citep{Lewis2000} to compute the linear theory mass fluctuations
power spectrum for the target cosmological parameter vector. This power
spectrum together with the expected growth history are then used to compute the
best scaling parameters, which in turn are used together with the pre-computed
projected correlation functions to predict the shear correlation functions
expected in the target cosmological scenario. We repeat this procedure for 8
independent chains with $10,000$ steps each and a burning phase of $200$ steps. 

To make the analysis feasible, we heavily optimised all numerical
codes involved. As a result, each step in the chain can be carried out in less
than $4$ seconds, including the linear-theory calculation and the
cosmology-rescaling algorithm.

\subsection{Cosmological constraints}
\label{sec:constraints}
\begin{figure} 
\includegraphics[width=\linewidth]{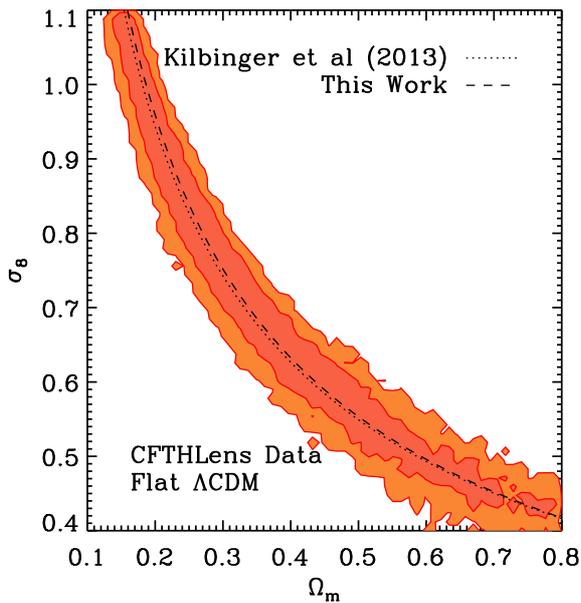}
\caption{
\label{fig:cosmological_constraints} 
Marginalised $68\%$ and $95\%$ confidence level intervals in the
$\Omega_m$-$\sigma_8$ plane derived from the shear correlation functions
measured by the {\cfh} survey. The solid blue line indicates the constraints
obtained for the same dataset by \citet{Kilbinger2013}, which follows the
degeneracy of constant $\sigma_8^{-1}\Omega_{\rm m}^{0.6}$.
}
\end{figure}

We now present the constraints estimated from the {\cfh} data. We focus on the
parameters $\sigma_8$ and $\Omega_{\rm m}$, as these contain most of the cosmological
information encoded in the shear correlation functions. The constraints on
the other 3 parameters of our model are rather weak. 

Fig. \ref{fig:cosmological_constraints} shows the two-dimensional marginalised
constraints in the $\Omega_{\rm m}$-$\sigma_8$ plane obtained from applying our model
to the measured shear correlation functions. The red and orange regions denote
the 68 and 96 per cent confidence levels. 

The constraints are elongated along the degeneracy that approximately
corresponds to a constant value of $\sigma_8^{-1} \Omega_{\rm m}^{0.6}$.  This is a
result of a combination of geometrical and growth effects. Our best-fit value
for this parameter combination is $\sigma_8^{-1}\Omega_{\rm m}^{0.6} =
0.801\pm0.028$, which is displayed as a long dashed line in Fig.
\ref{fig:cosmological_constraints}.

For comparison, the result of \cite{Kilbinger2013} is displayed as a dotted
line. These authors find $\sigma_8^{-1}\Omega_{\rm m}^{0.6} = 0.79\pm0.03$ using the
same dataset and a theoretical model based on an improved version of the
halofit code, which relies on the approximated fitting formula of
\cite{Eisenstein1998}.

We can see that our constraints are very similar those of
\citet{Kilbinger2013}. In particular, the statistical uncertainty is almost
identical between the two methods. The disagreement in the best fit combination
is only a small fraction of the statistical uncertainty in the constraints.
This validates our approach and confirms that our theoretical model is at
least, at the same level of accuracy as the state-of-the-art. We emphasise
though, that our formalism present many advantages over previous ones which
will be likely relevant for future data analyses, as it has been discussed
thorough our paper.

\begin{figure} 
\includegraphics[width=\linewidth]{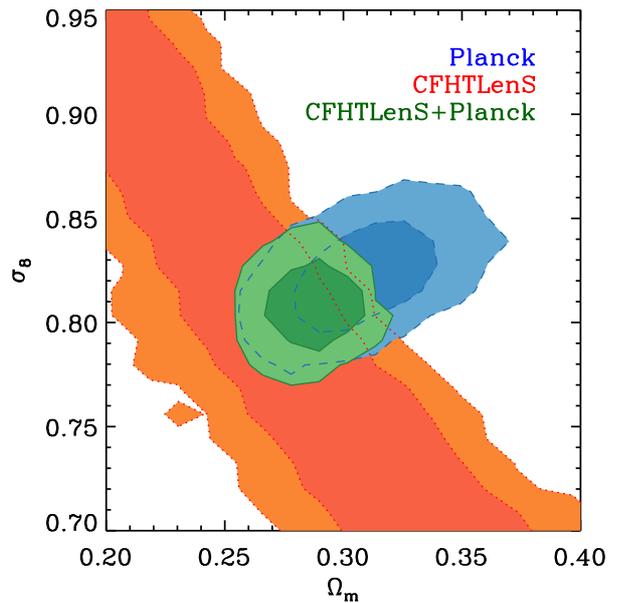}
\caption{
\label{fig:cosmological_constraints_with_Planck} 
Marginalised constraints in the $\Omega_m$-$\sigma_8$ plane. The dashed blue
contours corresponds to the measurements from the Planck data, whereas the red
contours shows the constraints from the {\cfh} shear correlation functions.
The green contours show the results when these two measurements are combined.
}
\end{figure}

In Fig.~\ref{fig:cosmological_constraints_with_Planck}, we combine our
constraints with those obtained from CMB measurements. Specifically, we use the
data described in \S\ref{sec:data:cmb} via publicly available
chains\footnote{\texttt{ http://pla.esac.esa.int/pla/aio/product-action?\\
COSMOLOGY.FILE\_ID=\\COM\_CosmoParams\_base\_planck\_lowl\_lowLike\_highL\_R1.10.tar.gz}},
which is displayed by the blue contours.

There is a mild tension between the two datasets considered. However, this is
not statistically significant. We thus combine the constraints by multiplying
the respective likelihoods. The addition of Planck data breaks the degeneracy
present in the {\cfh} constraints, resulting in constraints much tighter than
those from each of the methods individually.  The marginalised best fit
parameters we obtain are $\sigma_8 = 0.811 \pm 0.009$ and $\Omega_m = 0.290 \pm
0.011$, which is within the $1\sigma$ regions of each dataset.

\begin{figure} 
\includegraphics[width=\linewidth]{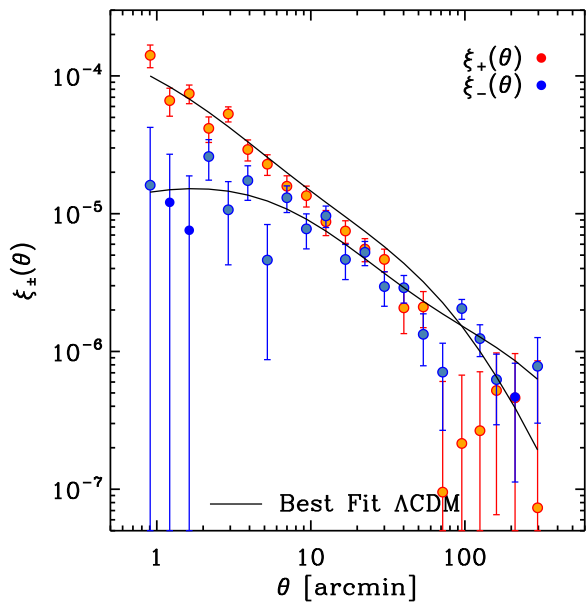}
\caption{
\label{fig:best_fit} 
The cosmic shear correlation functions, $\xi_+(\theta)$ (red circles) and
$\xi_-(\theta)$ (blue circles), as measured by the {\cfh} survey. The error
bars are given by the square root of the inverse of the respective covariance
matrix. The solid lines show the best-fitting flat $\Lambda$CDM model to the
shear correlation functions and to the CMB temperature anisotropies measured by
the Planck satellite and the polarization measured by the WMAP satellite. }
\end{figure}

Fig.~\ref{fig:best_fit} compares measured {\cfh} shear correlation functions to
those predicted by our method for these best fit parameters. We can see that
the model is indeed a very good description of the data at all scales and for
both $\xi_+$ and $\xi_-$, in particular if one keeps in mind the large
covariance among the points and that this is not necessarily the best fit to
the {\cfh} data alone. We also recall that in \S\ref{sec:nbody-scale}, we
carried out a direct $N$-body simulation with this specific set of parameters
and found that our theoretical model was accurate at the $<3-4\%$ level for
this target cosmological model.

\section{Conclusions}

We have developed and explored a new formalism for an accurate and flexible
cosmological exploitation of cosmic shear correlation functions. The core of
the method is an improved version of the cosmology-scaling algorithm originally
proposed by \cite{Angulo2010b}.

We have explicitly demonstrated the accuracy of our approach by comparing its
predictions with those derived from $N$-body simulations carried in extreme
cosmological scenarios. We showed that shear correlation function can be
predicted always with a $<10\%$ accuracy, which is much smaller than the
uncertainties in the data, and comparable with the theoretical uncertainties in
$N$-body simulations and in the procedure to combine mass correlation functions
into shear correlations.

The method can quickly provide, in about 4 seconds, predictions for
shear correlation functions. We exploited this by coupling the algorithm with an
MCMC sampler to obtain cosmological constraints on $\sigma_8$ and $\Omega_m$ from
{\cfh} data. The resulting constraints follow a degeneracy of constant
$\sigma_8^{-1}\Omega_{\rm m}^{0.6} = 0.801\pm0.028$, which can be broken with the
addition of CMB data. In this case, we obtain marginalised constraints $\sigma_8
= 0.811 \pm 0.009$ and $\Omega_m = 0.290 \pm 0.011$

Our results are very similar to published results, which rely in simpler (and
likely less accurate) input nonlinear correlation functions.  The current
agreement is due to the still relatively large statistical uncertainties
associated with the data. This situation, however, will change in the
foreseeable future with the arrival of several wide and deep weak lensing
surveys, which will demand greater accuracy and sophistication in the
cosmological analysis.

Here we have shown the  basic feasibility of our framework.  We foresee several
developments building upon the unique features our method provides. In
particular:

\begin{enumerate} 
\item[(1)] The high accuracy in the predictions will further increase naturally
as larger and higher-resolution $N$-body simulations arrive.  Furthermore,
simulations with different cosmological parameters can be combined to cover a
large range in parameter space, for which the scaling yields two-point and
higher-order correlation functions with a tight upper limit on their
uncertainties.

\item[(2)] The continuous progress in computer hardware and algorithms should
soon allow an incorporation of the large-scale structure correction into the
MCMC chains. Similarly, eventually it should be possible to directly handle the
full $N$-body outputs allowing, for instance, an on-the-fly light-cone
construction. 

\item[(3)] A full three-dimensionality in the predictions will be possible,
which will allow to dynamically incorporate selection functions, masks
appropriate of a given survey, and an accurate account for effects such as
source-lens clustering.

\item[(4)] The dark matter halo and subhalo population will be resolved at each
point of the cosmological parameter space. This could, in principle, enable a
flexible and realistic account of baryonic effects (such as mass redistribution inside halos due to gas cooling and outflows), e.g. via a combination
with galaxy formation models.
\end{enumerate}

It is for all these characteristics that we anticipate that our method could be
extremely useful to exploit the upcoming lensing data. A full realisation of
its potential could change the way in which cosmological data is analysed,
advancing considerably the state-of-the-art in terms of scope, sophistication
and realism.

\section*{Acknowledgements}

We thank Martin Kilbinger and Simon White for helpful comments.

We thank the CFHT staff and the members of the {\cfh} collaboration for
providing the {\cfh} shear correlation data products to the astronomical
community.

The {\cfh} data used in this publication is based on observations obtained with
MegaPrime/MegaCam, a joint project of CFHT and CEA/IRFU, at the
Canada-France-Hawaii Telescope (CFHT) which is operated by the National
Research Council (NRC) of Canada, the Institut National des Science de
l'Univers of the Centre National de la Recherche Scientifique (CNRS) of France,
and the University of Hawaii. This work is based in part on data products
produced at Terapix available at the Canadian Astronomy Data Centre as part of
the Canada-France-Hawaii Telescope Legacy Survey, a collaborative project of
NRC and CNRS.

{\it Author contributions}: R.A. and S.H. planned the project, wrote the paper
and interpreted the results. R.A. carried out the test simulations and the MS-I
into the future, devised an improved cosmology-scaling algorithm, computed the
3D and shear correlation functions, and performed the cosmological analysis.
S.H. derived the expressions for shear correlations from scaled mass
correlation functions, carried out the ray-tracing simulations, computed
projected mass and shear correlation functions, and performed systematics tests
on these.

\bibliography{database}

\label{lastpage} \end{document}